\documentclass[aps, prl, showpacs, reprint, showkeys, preprintnumbers,amsmath,amssymb,superscriptaddress,floatfix]{revtex4-1}
\usepackage[]{graphicx}      % Include figure files
\usepackage{bm}             	% bold math
\usepackage{times}			% Times font
\usepackage{tabularx}
\usepackage{color}

\begin{document}
\title{Exchange stiffness in ultrathin perpendicularly-magnetized CoFeB layers determined using spin wave spectroscopy}

\author{T. Devolder}
\email{thibaut.devolder@u-psud.fr}
\author{J.-V. Kim}  
\affiliation{Institut d'Electronique Fondamentale, CNRS, Univ. Paris-Sud, Universit\'e Paris-Saclay, 91405 Orsay, France}
\author{L. Nistor}
\author{R. Sousa}
\author{B. Rodmacq}
\author{B. Di\'eny}  
\affiliation{SPINTEC, UMR CEA/CNRS/UJF-Grenoble 1,  Grenoble-INP, INAC, Grenoble, F-38054, France}

\date{\today}                                           
%%%%%%%%%%%%%%%%%%%%%%%%%%%%%%%%%%%%%%%%
%
%       Abstract
%
%%%%%%%%%%%%%%%%%%%%%%%%%%%%%%%%%%%%%%%%
\begin{abstract}
We measure the frequencies of spin waves in nm-thick perpendicularly magnetized  FeCoB systems, and model the frequencies to deduce the exchange stiffness of this material in the ultrathin limit. For this, we embody the layers in magnetic tunnel junctions patterned into circular nanopillars of diameters ranging from 100 to 300 nm and we use magneto-resistance to determine which rf-current frequencies are efficient in populating the spin wave modes. Micromagnetic calculations indicate that the ultrathin nature of the layer and the large wave vectors used ensure that the spin wave frequencies are predominantly determined by the exchange stiffness, such that the number of modes in a given frequency window can be used to estimate the exchange. For 1 nm layers the experimental data are consistent with an exchange stiffness A= 20 pJ/m, which is slightly lower that its bulk counterpart. The thickness dependence of the exchange stiffness has strong implications for the numerous situations that involve ultrathin films hosting strong magnetization gradients, and the micromagnetic description thereof.
\end{abstract}

%\keywords{Exchange stiffness, spin waves, Magnetic Tunnel Junction, Perpendicular Magnetic Anisotropy, ferromagnetic resonance}

\maketitle

%%%%%%%%%%%%%%%%%%%%%%%%%%%%%%%%%%%%%%%%
%
%                Paper
%
%%%%%%%%%%%%%%%%%%%%%%%%%%%%%%%%%%%%%%%%

%\section{Introduction}
The exchange interaction is fundamental to magnetism as it underpins the existence of ordered spin states. It determines the energy scale of excitations such as spin waves and the length scales in topological spin structures such as domain walls and vortices. Its strength establishes the extent to which ordered magnetic states are robust against thermal fluctuations, by governing quantities such as the Curie temperature at which the ferromagnetic-paramagnetic phase transition takes place. Quantifying the exchange interaction is therefore important for both fundamental studies and technological applications in which magnetic materials are used.

For thin films with thicknesses greater than 10 nm, there exist a variety of experimental methods to determine the exchange stiffness, $A$. A number of techniques involve characterizing the spin wave dispersion, such as neutron scattering~\cite{alperin_observation_1966}, inelastic X-ray scattering~\cite{mulazzi_temperature_2008}, Brillouin light scattering~\cite{liu_exchange_1996}, and broadband ferromagnetic resonance using inductive methods~\cite{bilzer_study_2006}.  Other methods include the direct use of Bloch's law~\cite{nembach_linear_2015}. For these thick systems, it is generally found that $A$ is similar in magnitude to its bulk counterpart. However, since the exchange interaction involves the orbital overlap of the constituent magnetic atoms, the stiffness $A$ should be affected by the reduction in coordination number. This is a well established phenomenon in alloys \cite{antoniak_composition_2010, eyrich_effects_2014} but should also be important in ultrathin films in which a larger proportion of the material is exposed to surfaces and interfaces. This has indeed been found for iron: an exchange stiffness of $A= 2 \pm 0.4$ pJ/m has been determined for an epitaxial monolayer of Fe sandwiched between Ir and Pd~\cite{romming_field-dependent_2015} while a value of $A = 11.9 \pm 4$ pJ/m has been found for monolayer superlattices of Fe/Pt~\cite{antoniak_composition_2010, okamoto_chemical-order-dependent_2002}, which are much lower than the iron bulk value of $A = 20$ pJ/m~\cite{cochran_light_1994}.

For ultrathin films, scattering methods become less useful because they are not sufficiently sensitive to detect spin waves. Furthermore, techniques based on the Bloch law require extreme care as the thermal dependence of the magnetization is very sensitive to the range of the interactions and to the dimensionality of the system~\cite{mermin_absence_1966, krey_significance_2004, nembach_linear_2015}. As such, many attempts to date have relied on spin textures with large magnetization gradients which require some assumptions on the form of the micromagnetic states involved~\cite{romming_field-dependent_2015, okamoto_chemical-order-dependent_2002}. On the other hand in magnetoresistive multilayers such as spin valve or magnetic tunnel junction (MTJ) nanopillars, it is possible to obtain signatures of the magnetization dynamics on lengths scales determined by the pillar geometry. For instance, estimates of $A$ have been obtained in spin-transfer torque magnetic random access memories (STT-MRAM) by measuring the thermal stability and magnetization reversal involving spatially nonuniform processes~\cite{sato_CoFeB_2012}, and by characterizing the thermal noise of confined spin wave eigenmodes~\cite{devolder_spin-torque_2011}. For these in-plane magnetized systems, it has been shown that a reduction of $A$ with thickness from 20 pJ/m (bulk Fe) and 30 pJ/m (bulk Co) to 23 pJ/m for \cite{devolder_spin-torque_2011}  2 nm of Fe$_{40}$Co$_{40}$B$_{20}$ and 19 pJ/m for \cite{sato_CoFeB_2012}  1.6 nm of Fe$_{56}$Co$_{19}$B$_{25}$.

However, an open question remains on how the exchange interaction evolves in ultrathin films with perpendicular magnetic anisotropy (PMA), such as nm-thick (Co,Fe) alloys in contact with heavy-metal underlayers and metal oxides, which underpin many studies on spin-orbit torques and interface-driven chiral interactions at present. In this article, we describe the exchange stiffness and its thickness dependence in ultrathin CoFeB PMA layers sandwiched between Ta and MgO, which have been deduced from spin wave spectroscopy in circular MTJ nanopillars. For 1 nm thick layers the spin wave frequencies are consistent with an exchange stiffness of $A=20\pm2$ pJ/m, which is found to be slightly lower than the bulk value of 27.5 pJ/m. The exchange stiffness is not found to decrease as dramatically as the magnetization with decreasing film thicknesses towards atomic dimensions.
%%%%%%%%%%%%%%%%%%%%%%%%%%%%
%%%%%%%%%%%%%%%%%%%%%%%%%%%%
%%%%%%%%%%%%%%%%%%%%%%%%%%%%

%%%%%%%%%%%%%%%%%%%%%%%%%%%%
%%%%%%%%%%%%%%%%%%%%%%%%%%%%
%%
%	Figure
%%
%
\begin{figure}
\includegraphics[width=9 cm]{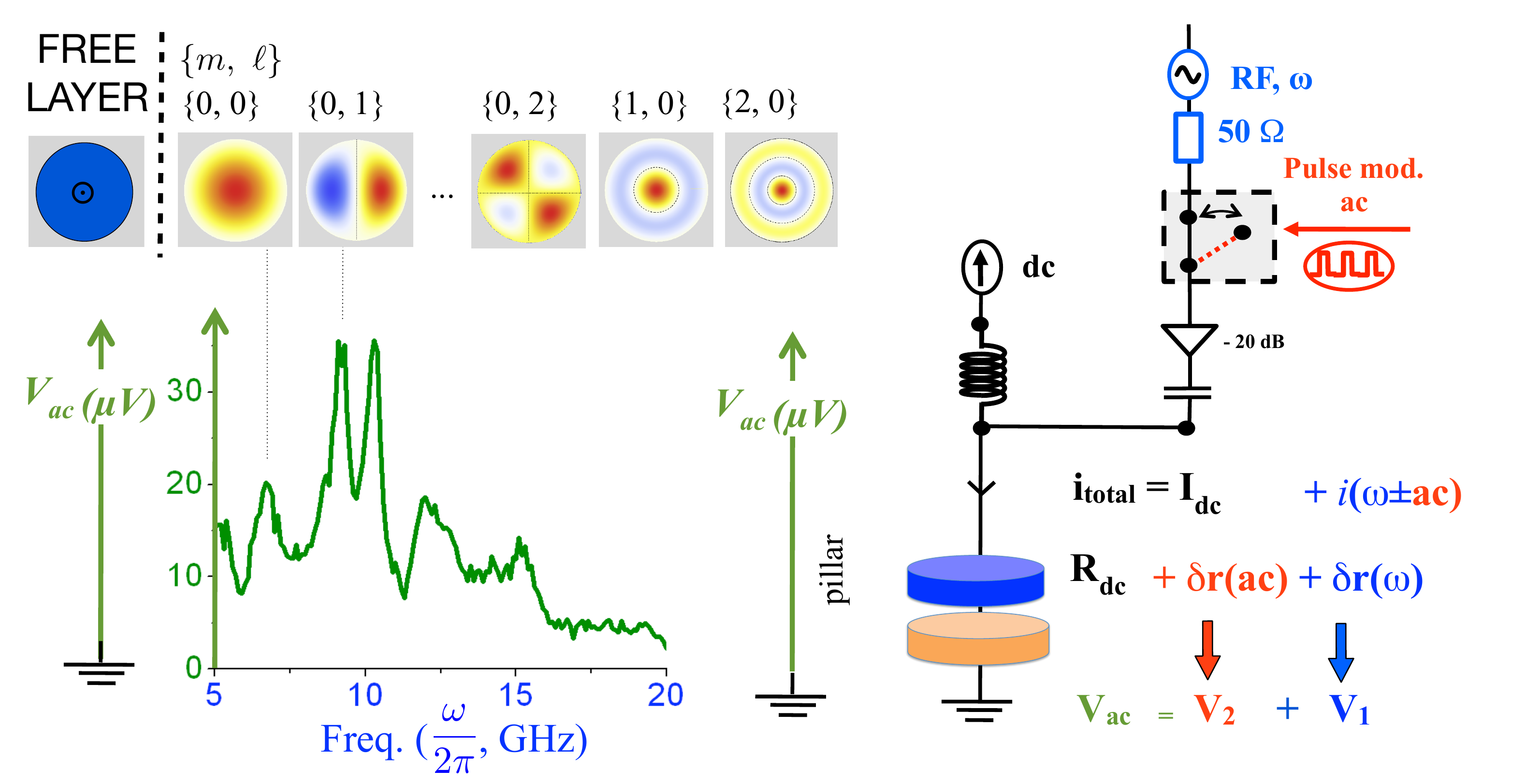}
\caption{Illustration of the method and sketch of the set-up. Right: the pillar is biased with a dc voltage and an RF voltage undergoing an on/off modulation at an ac frequency. The pillar resistance oscillations create an ac voltage drop, which is peaked (bottom left) whenever the applied rf frequency hits a spin wave frequency. Top left: examples of the $x$ component of the dynamic magnetization of some of the eigenmodes of a uniformly magnetized free layer disk (blue) according to the analytical frameworks \cite{klein_ferromagnetic_2008}. The superimposed dashed lines are the nodal lines.}
\label{setup}
\end{figure}

\begin{figure}
\includegraphics[width=8.5cm]{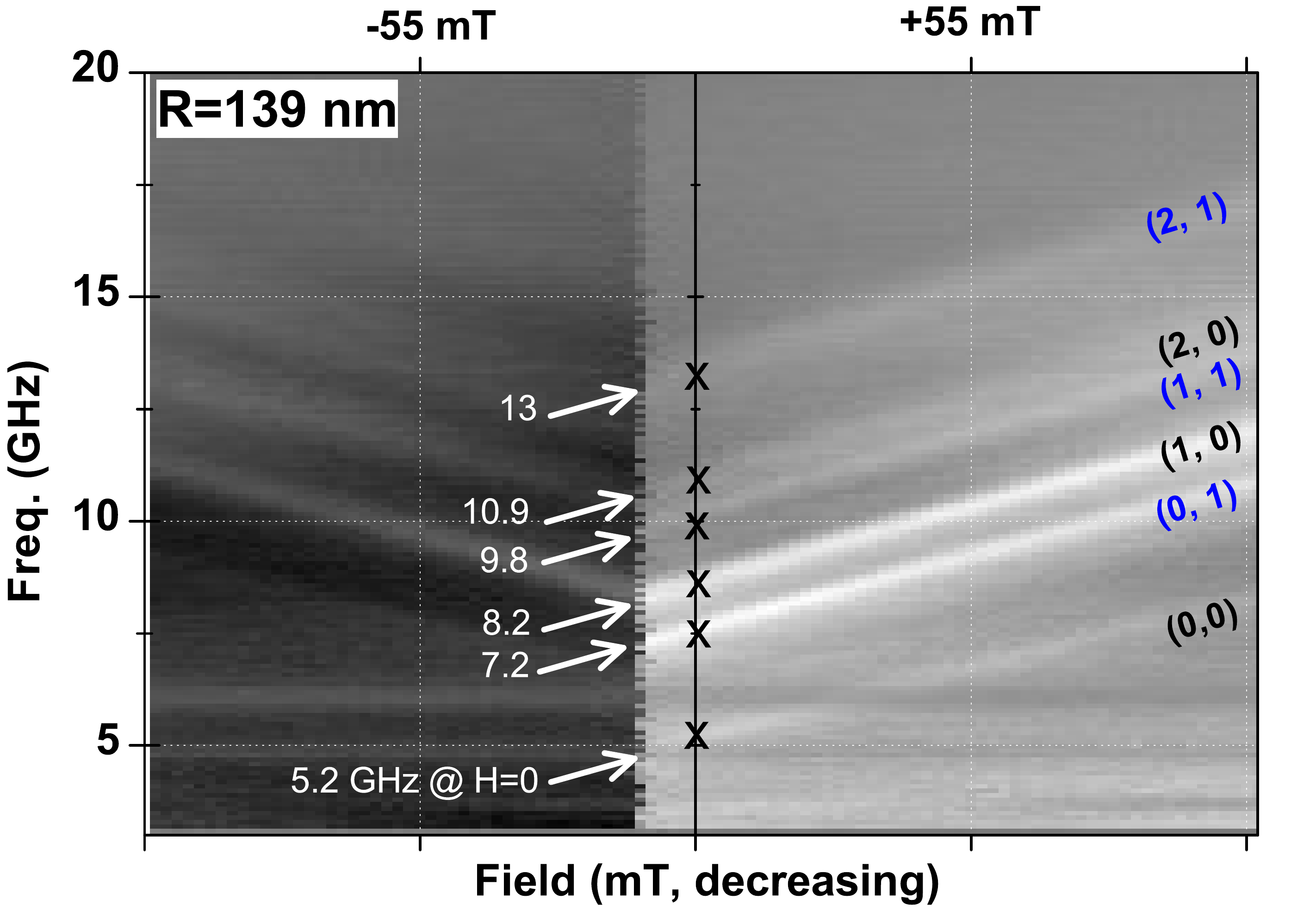}
\caption{Rectified voltage versus field and frequency for a nanopillar of radius 139 nm and a free layer of thickness 10 \r{A} biased with $V_{dc}=-1~\textrm{mV}$.  The field is decreased from 110 mT (AP state) to -110 mT. The pixelized line of color change at -13 mT is the switching to the P state. The superimposed labels recall the proposed mode indexation. The quoted frequencies correspond to the zero field point.}
\label{STTFMRA4}
\end{figure}

%\section{Samples}
Our samples are MTJs with PMA. The stack composition is: buffer/Ta (5)/ Co$_{60}$Fe$_{20}$B$_{20}$ (1)  (free layer)/ MgO / Co$_{60}$Fe$_{20}$B$_{20}$ (1.6)/ [Tb (0.4)/Co (0.5)]$_{\times 20}$ /Ta (3)/Cu (3)/Ru (7), where the numbers in parentheses represent thicknesses in nm. The Tb/Co reference multilayer is a ferrimagnetic alloy, whose moment is nearly compensated (magnetization is 75~kA/m) to minimize the stray field acting on the free layer. The Tb/Co reference multilayer has a coercivity of 0.6~Tesla, and an effective anisotropy above 1 Tesla. This large anisotropy ensures that the spin waves in the reference layer have frequencies above 30~GHz, which can be neglected in the ensuing analysis since the free layer modes of interest lie in the range of 0-20 GHz (Fig.~\ref{STTFMRA4}). The free layer is crystallized \cite{lee_giant_2006} during an annealing step that is accompanied by a partial diffusion of the B atoms towards the Ta layer, leaving a free layer whose composition lies between Co$_{75}$Fe$_{25}$ and Co$_{60}$Fe$_{20}$B$_{20}$. The B interstitial atoms are known not to affect substantially the exchange stiffness which is $A=27.5$ pJ/m in the bulk \cite{bilzer_study_2006}. The layers were not found to be ferromagnetic at room temperature for thickness below 0.5 nm, and exhibited PMA for thickness up to 1.2 nm. We do not make any assumptions on the presence of any magnetically ``dead'' layers, so the free layer magnetization is defined as the areal moment divided by the nominal thickness. The magnetizations were measured by vibrating sample magnetometry and were found to increase linearly from $M_S = 0.6$ MA/m for 1 nm thick layers to a plateau with a bulk value (1.4 MA/m) for thicknesses above 1.5 nm. The tunnel magnetoresistance is typically $\eta_{tmr} = 50\%$. The stack resistance-area product is $\textrm{RA}=14~ \Omega.\mu \textrm{m}^2$.

The PMA MTJs were patterned into circular pillars with nominal diameters $2a$ of 100, 200, and 300 nm. For the set of devices studied, the mean device resistances were found to be correlated with their nominal size, but the slight dispersion in the values observed suggests a dispersion in the radii of $\pm 15 \textrm{~nm}$. We have evidence (not shown) that the junctions are largely circular, so in the remainder of this study we will assume an exact junction radius defined by $a = \sqrt{ RA / (\pi R_p)}$. This leads to typical resistances of 200 $\Omega$ (a=150 nm) to 1800 $\Omega$ (a= 50 nm) in the parallel (P) state.

\begin{figure}
\includegraphics[width=8.5cm]{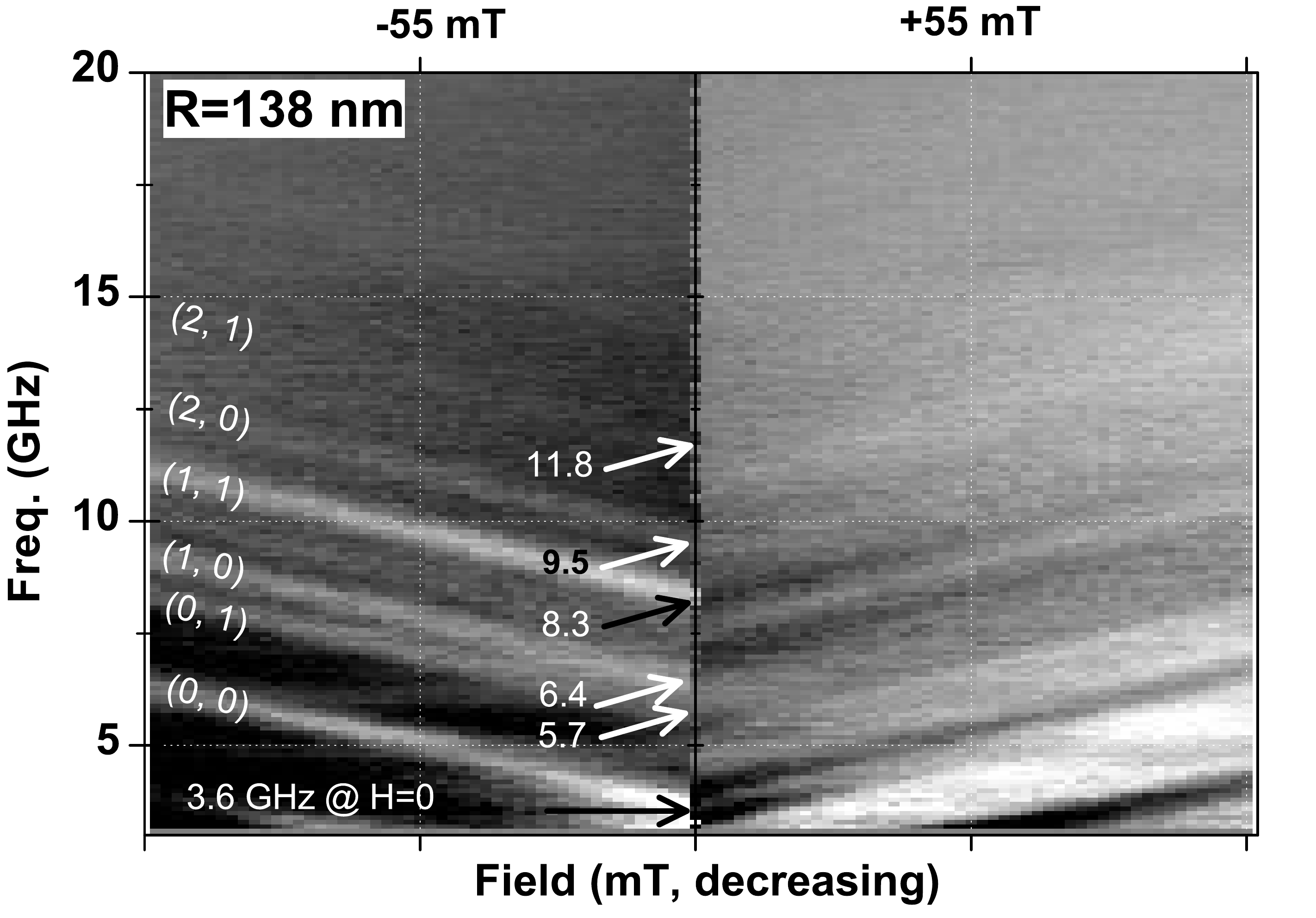} %Fig2_KdB4
\caption{Rectified voltage versus field and frequency for a nanopillar of radius 138 nm and a free layer of thickness 10.5 \r{A} biased with $V_{dc}=+1~\textrm{mV}$. The field is decreased from 110 mT (AP state) to -110 mT. The pixelized line of color change at -1 mT is the switching to the P state. The superimposed labels recall the proposed mode indexation. The quoted frequencies correspond to the zero field point. }
\label{Kserie}
\end{figure}

%%%%%%%%%%%%%%%%%%%%%%%%%%%%
%\section{Set-up and sensitivity function}
The pillars are characterized in an STT-FMR-like \cite{tulapurkar_spin-torque_2005} set-up (Fig.~\ref{setup}). The objective is to reveal the frequency versus field dispersion laws of the spin waves of the free layer. The device is biased using a dc source supplying $V_\textrm{dc}= \pm 1~ \textrm{mV}$ and fed with an RF voltage supplying  $V_{rf} \approx 350\textrm{~mV}_{pp}$ at a variable frequency $3 \leq \omega / (2\pi) \leq 20~\textrm{GHz}$. The RF voltage amplitude is pulse-modulated at an ac frequency $f_{ac} = 50~\textrm{kHz}$ (Fig.~\ref{setup}). The spin waves are populated by the RF torques, i.e., the STT, along with the current-induced fields that are generated by the microwave circuitry used to probe the device. A proper demodulation of the ac voltage across the device yields a signal containing magnetic susceptibility information. For each applied field, the signal exhibits marked extrema at well defined frequencies (Fig.~\ref{setup}).
To check that the signal extrema are not artefacts due to impedance mismatches and the related standing waves in the circuitry, we vary the applied magnetic field. This is for instance done in Fig. \ref{STTFMRA4} and \ref{Kserie} for large MTJs with 10 \r{A} and 10.5 \r{A} thick free layers. Several points are worth noticing in these figures. \\
(i) The spin wave dispersion relations appear as V-shaped branches, with a slope inversion when the magnetization of the FeCoB free layer switches. This slope inversion confirms that the observed spin waves are hosted by the free layer. The peak frequency linewidths $\Delta \omega$ are consistent with a free layer damping of $\Delta \omega / (2\omega) = 0.03-0.04$, depending on device. For a given device diameter, the spin waves gradually move to lower frequencies as the free layer thickness is increased and the effective anisotropy is lowered (Fig.~\ref{Kserie}). The spin wave branches are linear for samples showing PMA. The spin wave branches bend in a complex manner (not shown) for thicker samples when the magnetization tilts towards the in-plane direction.  
When reducing the device diameters, this increases the lateral confinement of the spin waves which increase of the exchange contribution to their frequencies. Consequently, the frequency spacing between the branches increases (see supplementary document).\\
(ii) In addition, the switching of the free layer results also in a drastic change of the signal amplitude, that can be noticed as a left-side versus right-side contrast in Fig.~\ref{STTFMRA4}. While the modes are clearly identified for one field polarity, their identification is systematically much more difficult for the other field polarity. This drastic dependence of the demodulated signal on the free layer orientation is informative on the nature of the mechanism that populates the observed spin waves. Let us thus discuss the origin and the amplitude of our signal. \\

%%%%%%%%%%%%%%%%%%%%%%%%%%%%
%%%%%%%%%%%%%%%%%%%%%%%%%%%%
%%
%	Figure
%%
%

The demodulated signal can contain two ac components $V_1$ and $V_2$ (Fig.~\ref{setup}) of different physical origins. The first expected component is the standard STT-FMR signal: the rf current rectifies any synchronous oscillation of the resistance. However because of the cylindrical symmetry of our configuration, the magnetization precession associated with a spin wave is supposed not to make the device resistance oscillate at the precession frequency. The STT-FMR signal $V_1$ should thus vanish \cite{naletov_identification_2011}.
The second signal ($V_2$) is a much larger signal related to the decrease of the time-averaged magnetization due to the spin wave populations. Indeed when the RF torques are applied the spatially and temporally averaged magnetization $|\langle M_z \rangle|$ of the free layer is less than its value $M_S$ in the saturated states. The correlated change of resistance is revealed by the small dc current passing through the sample, i.e. 
\begin{equation} 
V_2 \propto I_{dc} \times (\langle M_z\rangle-M_S)~. 
\end{equation}
Note that $V_2$ changes sign with the sign of the bias current $I_{dc}$ and with free layer switching, in line with our finding (see the left-right contrast is Fig.~\ref{STTFMRA4}). The amplitude of $V_2$ scales with the rf Oersted field (see supplementary document), hence it decreases as the device diameter. \\

%%
%	Figure
%%
%

Let us now describe the nature of the observed spin waves. The eigenmodes of perpendicularly magnetized circular disks are well understood \cite{klein_ferromagnetic_2008, naletov_identification_2011, munira_calculation_2015}. They can be referred to by their radial index $m \in \mathbb{N}$  where $m$ is the number of nodes along the radial axis and $\ell \in \mathbb{Z}$ is the winding number which describes the number of magnetization turns along any path encompassing the disk center. The modes $\{m, 0\}$ are the purely radial modes of symmetries $\{$ \setlength{\unitlength}{0.1mm}
\begin{picture}(40, 40)
\put(0,10){\circle{40}}
\end{picture}
\begin{picture}(40, 40)
\put(0,10){\circle{15}}
\put(0,10){\circle{40}}
\end{picture}      
\begin{picture}(40, 40)
\put(0,10){\circle{10}}
\put(0,10){\circle{25}}
\put(0,10){\circle{40}}
\end{picture}       
\begin{picture}(40, 40)
\put(0,10){\circle{10}}
\put(0,10){\circle{20}}
\put(0,10){\circle{30}}
\put(0,10){\circle{40}}
\end{picture}...$\}$ where the lines denote the nodes that separate regions of dynamical magnetizations with opposite signs. The azimuthal index $\{ \ell=1,~2,... \}$ adds nodal lines $\{ {\ominus},~\oplus,...\}$ along the sample diameters (see some examples in Fig.~\ref{setup}). In isolated perfectly circular disks with PMA, the  $\{m, \ell\}$ and $\{m, -\ell\}$ modes are degenerate in the absence of magneto-static interactions \cite{munira_calculation_2015}. While these conditions are not strictly met (see suppl. document) we do not expect to be able to lift the degeneracy and we will thus sort our modes by considering $\ell \geq 0$ only.

The different spatial profiles of the spin wave modes (Fig.~\ref{setup}) confers to them different susceptibilities to the RF torques present in our configuration. The rf current flowing through the electrodes generates a small quasi-uniform rf field $h_x$ on the device. This field can excite the purely radial $\{m, 0\}$ modes but the excitation efficiency decrease with $m$. This parasitic pumping field $h_x$ is rather uniform at the scale of the device such that it should not excite $\ell \geq 1$ modes. 
Conversely the rf Oersted field has a $|\ell| = 1$ symmetry and thus excites predominantly the $\{m, 1\}$ modes. It has a strong radial gradient which provides a priori an excitation route for all values of $m$. Since the dynamic Oersted field provides the largest torque in our configuration \cite{naletov_identification_2011}, we expect the strongest signal from the $\{m, 1\}$ modes. We have not identified any mechanism to excite the $\ell \geq 2$ modes in our experimental configuration. \\

The frequencies of the spin waves can be predicted semi-quantitatively using the model of ref.~\cite{naletov_identification_2011, munira_calculation_2015}. Writing the wavevectors such that $a k_{m, \ell}$ is the $m^{\textrm{th}}$ zero of the $\ell^{\textrm{th}}$ Bessel's function \cite{klein_ferromagnetic_2008}, the spin wave frequencies can be expressed as 
\begin{equation}
\omega_m^2 / \gamma_0^2= H_1 H_2  
\label{DispersionRelation}
\end{equation}
with the two stiffness fields being :
\begin{equation}
H_1 = H_z + \frac{2 K_s}{\mu_0 M_S t} - N_{m, \ell} M_S + \frac{2A k_{m, \ell}^2}{\mu_0 M_S}
\label{H1}
\end{equation}
and 
\begin{equation}
H_2 = H_1 + M_S (1-\frac{1-e^{k_{m, \ell} L}}{k_{m, \ell} L})
\label{eqH2}
\end{equation}
where $K_s$ is the interface anisotropy. While the last term in Eq.~\ref{eqH2} is the same as in infinite films \cite{kalinikos_theory_1986}, the factor $N_{m, \ell}$ is a mode dependent demagnetizing factor that accounts mainly for the reduction of the demagnetizing field in the out-of-plane direction as the disk radius is reduced. The demagnetizing terms for the purely radial modes $\{m, 0\}$ can be calculated analytically \cite{ross_standing_2014} but micromagnetics is unavoidable for the azimutal modes.

\begin{figure}
\includegraphics[width=8 cm]{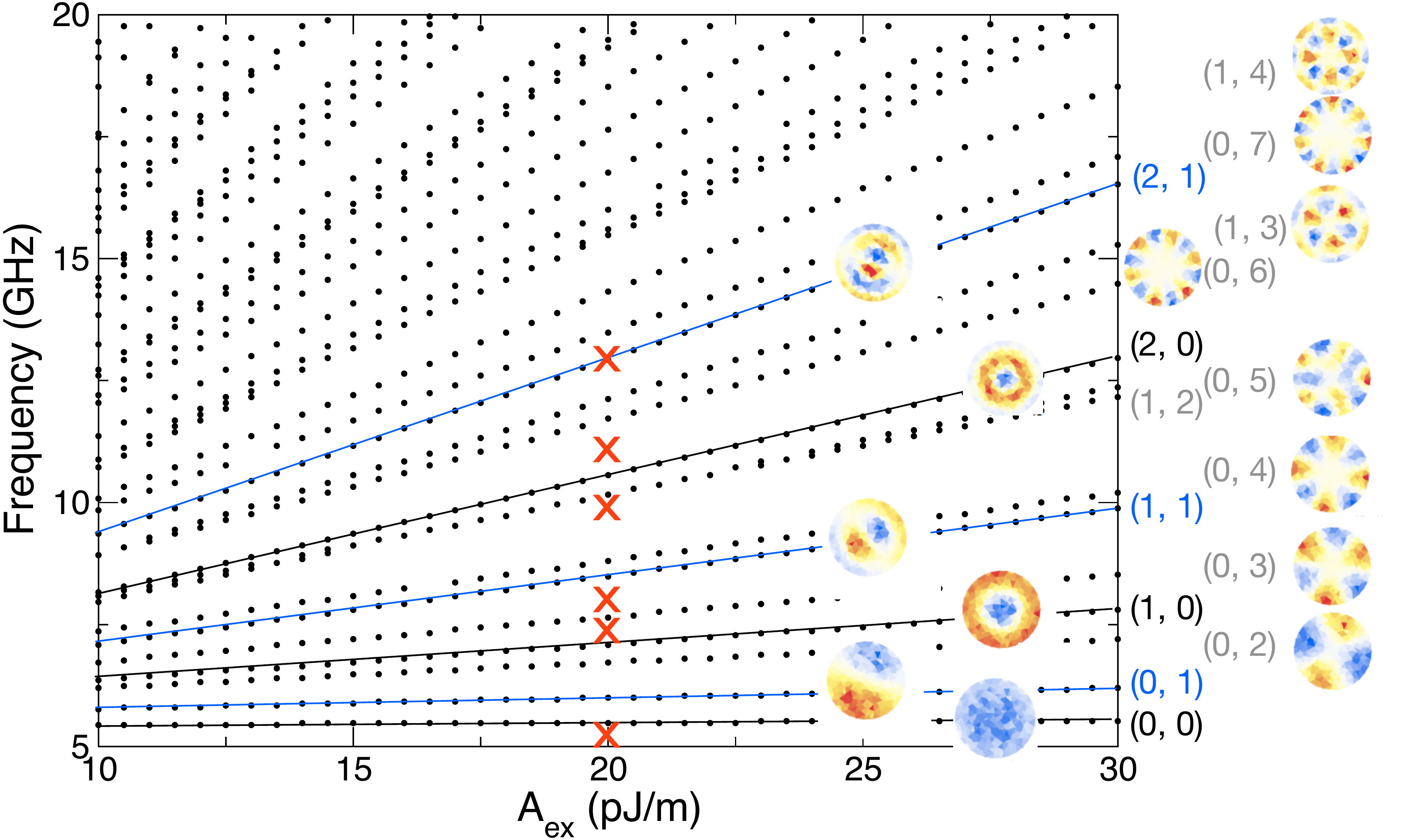}
\caption{Dependence of the micromagnetically calculated eigenmode frequencies (dots and lines as guide to the eye) versus exchange stiffness for circular disks of radius 138 nm, thickness 10 \r{A}, anisotropy 0.28 MJ/m$^3$ and magnetization 0.6 MA/m mimicking the sample on which Fig.~\ref{STTFMRA4} was recorded. The colored wheels illustrate the spatial profiles of each simulated mode. The red crosses are the experimentally detected frequencies.}
\label{micromagnetics}
\end{figure}

We have thus used micromagnetic simulations to take into account the magnetostatic interactions in an exact manner. Compared to the trial  eigenmode profiles used in the analytical frameworks (Fig.~\ref{setup} and refs. \cite{klein_ferromagnetic_2008, naletov_identification_2011, munira_calculation_2015}), the micromagnetics mode profiles (Fig.~\ref{micromagnetics}) are distorted and show a trend towards more localization near the sample edges. Linked with this localization, the influence of exchange interaction on the eigenmode frequencies (Eq.~\ref{H1}) is altered and the spacing between the frequencies does not appear to increase proportionally to the exchange stiffness (notice for instance the weak dependence on exchange of the spacing between the modes (0, 0) and (0, 1) in Fig.~\ref{micromagnetics}). As a result the sole mode-to-mode frequency distances cannot be used as a measure of $A/M_s$, contrary to what the (approximate) Eq.~\ref{DispersionRelation} would imply. \\
The deficiencies of Eq.~\ref{H1}-\ref{eqH2} do however not impede the mode indexation. Indeed even in full micromagnetics (Fig.~\ref{micromagnetics}) the frequencies of the modes are still ordered as they would be in the analytical frameworks according to their exchange terms $ \frac{2A k_{m, \ell}^2}{\mu_0 M_S}$. As such the modes always appear in the same order. Keeping only the modes with $\ell \leq 1$, the order is predicted to be $\omega_{(0, 0)} < \omega_{(0, 1)} < \omega_{(1,   0)} < \omega_{(1, 1)} < \omega_{(2, 0)} < \omega_{(2, 1)}$. Looking back at the experimental results, the lowest frequency mode [labeled (0,0)] led statistically to a low rectified signal. The two next modes [labeled (0,1) and (1,0)] led generally to the largest response. The weak (0,0) mode is logically the response to the (weak) rf field $h_x$ and the intense (0,1) mode is the response to the (larger) rf Oersted field. The next mode, labeled (1,0) and of radial character is more intense than anticipated, as if the pinning conditions at the device edge were much more relaxed compared to expectations. \\

%%
%	Figure
%%
%

We have conducted this indexation for several device sizes with 1 nm thick free layers. In the frequency window [5, 15 GHz], we usually found 6 modes for devices of radii in the range of 140 nm, 4 modes for radii 100 nm and 2 modes for radii 50 nm (see supplementary material). The experimental frequencies are spread from sample to sample, which indicates that the pinning conditions at the edges differ from device to device, probably as a result the patterning process. As shown in Fig.~\ref{micromagnetics} the observed frequencies do not generally match with the one predicted by micromagnetics. However with the measured areal moments and the nominal thicknesses, the exchange stiffness leading to 6 observable (i.e. with $\ell \leq 1$) modes in our measurement window was found to be $20\pm2$ pJ/m for 10 \r{A} thick free layer. Despite our sizable error bar, we can conclude that the exchange stiffness in the ultrathin limit is lower than that of the bulk state (27.5 pJ/m). However, its decay with the thickness seems much slower than the magnetization and does not follow the $A\propto M_S^2$ trend generally observed in bulk systems \cite{mulazzi_temperature_2008}. \\

%\section{Conclusion}
In summary, we have measured the frequencies of spin waves of large wavevectors in nm-thick perpendicularly magnetized CoFeB systems. The measurement method relies on magneto-resistance to enable the spectroscopy of rf-current populated spin waves in nanopillars of deep submicron size. The ultrathin nature of the films and the large wavevectors used ensure that the spin wave frequencies are essentially determined by the intralayer exchange stiffness. A proper mode indexation for several sizes of junctions allows to deduce that the exchange stiffness of the CoFeB free layer is around A= 20 pJ/m, which is slightly less that its bulk counterpart. The decrease in magnetization at low thickness is much more pronounced, such that the exchange length should be longer in the ultrathin films than in their bulk counterpart. This unexpected thickness evolution of the exchange stiffness has strong implications for the vast class of physical problems involving strong magnetization gradients in ultrathin films.

\begin{acknowledgements}
We thank Felipe Garcia-Sanchez and Gr\'egoire de Loubens for fruitful discussions. This work was partially supported by the French National Research Agency (ANR) under Contract No. ANR-11-BS10-0003 (NanoSWITI).
\end{acknowledgements}

\end{document}